\documentclass[twocolumn,superscriptaddress,aps,showpacs,floatfix,10pt, prl]{revtex4-1}
\usepackage[T1]{fontenc}
\usepackage{amssymb,amsmath}
\usepackage{graphicx}
\usepackage{color}
\usepackage{bbm}
\usepackage{bbold}
\usepackage{soul}
\usepackage[dvipsnames]{xcolor}
\usepackage{ulem}
\newcommand{\ah}{\hat{a}}

\newcommand{\expect}[1]{\langle #1 \rangle}

\begin{document}
\title{Dynamical blockade in a single mode bosonic system}
\author{Sanjib Ghosh}
\email{sanjib.ghosh@ntu.edu.sg}
\affiliation{School of Physical and Mathematical Sciences, Nanyang Technological University 637371, Singapore}
\author{Timothy C. H. Liew}
\email{timothyliew@ntu.edu.sg}
\affiliation{School of Physical and Mathematical Sciences, Nanyang Technological University 637371, Singapore}
\begin{abstract}
We introduce a dynamical blockade phenomenon occurring in a nonlinear bosonic mode induced by a combination of continuous and pulsed excitations. We find that the underlying mechanism for the blockade is general, enhancing antibunching in the strongly nonlinear regime and inducing it in the weakly nonlinear regime, without fine-tuning the system parameters.
Moreover, this mechanism shows advantages over existing blockade mechanisms and is suitable for implementation in a wide variety of systems due to its simplicity and universality. 
\end{abstract}
\maketitle
Photon blockade is a nonlinear optical effect that suppresses multiple-photon occupancy in a quantum mode favouring the single photon state~\cite{Imamoglu97}. Strong photon blockade is a natural source for single photons, which are essential for many rising technologies~\cite{Lodahl15}, such as quantum communication~\cite{Kimble08,Sangouard11}, computation~\cite{Knill01} and cryptography~\cite{Scarani09}. Accessing the regime of photon blockade is also a prerequisite for realizing quantum many-body phenomena, e.g., the fractional quantum Hall effect~\cite{Umucallar12}, the superfluid to Mott insulator transition~\cite{Hartmann06,Angelakis07,Greentree06} and the strongly correlated Tonks-Girardeau gas~\cite{Carusotto09} of photons.

While photon blockade has been realized in a variety of physical systems, they operate with  diverse mechanisms and methods in different regimes of the system parameters. Conventional photon blockade relies on the anharmonic energy spectra of multiple photons in a nonlinear cavity~\cite{Imamoglu97}. Naturally, this mechanism is inefficient in the weakly nonlinear regime where the corresponding spectral anharmonicity is smaller than the linewidth.  Consequently, the search for strong nonlinearity was the paradigm in this field, and it took different routes to enhance nonlinearity, e.g., by coupling photonic modes to single atoms \cite{Birnbaum05, Dayan08}, quantum dots \cite{Faraon08}, superconducting qubits~\cite{Lang11}, Rydberg atoms~\cite{Ningyuan18}, mechanical resonators~\cite{Lemonde16,Rabl11}, 2D materials~\cite{Ryou18} and doubly resonant nanocavities~\cite{Majumdar13,Gerace14}. Exciton-polaritons in semiconducting microcavities were also considered for inducing polariton blockade~\cite{Verger06} which was observed in recent experiments~\cite{Munoz-Matutano19, Delteil19} with however 
a limited antibunching due to limited nonlinearity. The regime of strong nonlinearity was recently accessed in exciton-polariton systems~\cite{Sun17, Rosenberg18, Togan18}, where the blockade physics would be exciting to study. 

Alternatively, an interference effect between a pair of coupled quantum modes can induce unconventional photon blockade in the weakly nonlinear regime \cite{Liew10, Bamba11, Bamba11APL, Lemonde14, Flayac17}, which was realized in recent experiments \cite{Vaneph18, Snijders18}. However, the emission-correlation in the unconventional blockade rapidly oscillates in time~\cite{Liew10}, requiring high time resolution to observe, as well as making it unsuitable for many applications. Other blockade mechanisms were proposed, based on gain media~\cite{Ghosh18}, parametric interactions~\cite{Kyriienko14,Sarma17}, and time-modulated driving fields~\cite{Kryuchkyan16}. Also, proposals to enhance the unconventional blockade have been based on phase dependent tunnelling~\cite{Shen15PRA}, multiple optomechanical modes~\cite{Sarma18}, and continuous bimodal driving~\cite{Shen17,Shen18}.

Here, we introduce a mechanism for photon blockade that can be dynamically induced universally in all regimes of nonlinearity. In our scheme, we resonantly apply a combination of both continuous and pulsed excitations to a nonlinear  bosonic mode. While either of the continuous or pulsed excitations individually induces a conventional blockade, their combined effect dramatically alters the scenario with a much stronger photon blockade in certain periodic time windows. The scheme is conceptually simple, because the system involves only a single mode driven by resonant optical fields that are routinely used in experiments (e.g., Ref.~\cite{Adiyatullin17}). The underlying mechanism is very general and can be applied to any nonlinear bosonic system. Moreover, the induced dynamical blockade has advantages over the existing blockade mechanisms, e.g., it shows no rapid oscillations in the unequal time correlation function like the unconventional blockade shows, and presents improved single photon statistics compared to that of the conventional blockade in its optimal operating configuration. Thus, the dynamical blockade can be used in preexisting single photon devices to improve their emission efficiency (brightness) and single photon statistics, while allowing other systems with weaker nonlinearity to reach the blockade regime.

Our theoretical description of the considered bosonic nonlinear mode driven by resonant optical fields is based on the quantum master equation. By analysing the system, we find the essential ingredients for the dynamical blockade to occur and identify the underlying mechanism. We present comprehensive numerical evidence for the phenomenon in different regimes of the mode parameters.

\textbf{The model:--} Let us consider a driven-dissipative Kerr nonlinear quantum mode represented by the Hamiltonian:
\begin{eqnarray}
\hat{H} = E\ah^\dagger \ah + \alpha\ah^\dagger \ah^\dagger \ah \ah + P(t)\,  \ah^\dagger  + P^*(t)\ah 
\end{eqnarray}
where $\ah^\dagger$ ($\ah$) is the creation (annihilation) operator, $E$ is the mode energy, $\alpha$ is the strength of nonlinearity and $P(t)$ represents the envelope of a coherent driving field (laser). It is implicit that we operate in the frame rotating at the laser frequency, such that E is the mode energy relative to the laser energy. The quantum master equation describing the dynamics of an observable $\hat{\mathcal{O}}$ is given by,
\begin{eqnarray}
i\hbar\expect{ \dot{ \hat{ \mathcal{O} } }} = \expect{[ \hat{\mathcal{O}},\hat{H}]} + i\frac{\gamma}{2}\expect{2\ah^\dagger \hat{\mathcal{O}}\ah - \ah^\dagger \ah  \hat{\mathcal{O}} - \hat{\mathcal{O}} \ah^\dagger \ah }
\end{eqnarray}
where $\gamma/\hbar$ is the decay rate of the mode.
As a measure of antibunching, we consider the second order correlation function:
\begin{eqnarray}
g_2(t,t') = \frac{\expect{\ah(t)^\dagger \ah(t')^\dagger \ah(t')  \ah(t)} }{\expect{\ah(t)^\dagger \ah(t)}  \expect{\ah(t')^\dagger \ah(t')}}
\end{eqnarray}
that represents the correlation between emission at times $t$ and $t'$. For ideal single photon emission, a vanishing equal time correlation function $g_2(t,t)$ is required. The dynamics of the equal time correlation function can be obtained from the master equation (see Ref.~\cite{SM}): 
\begin{eqnarray}
 \dot{g}_2(t,t) = \frac{4P(t)}{ \hbar} f(t)
 \label{g2Dot}
\end{eqnarray}
where the function $f(t) = \left( g_2(t,t) n\, \text{Im}[\psi] -  \text{Im}[C] \right)/ n^2  $ with occupation number $n=\expect{\ah^\dagger\ah}$, the mean field wave-function $\psi = \expect{\ah}$ and $C=\expect{\ah^\dagger \ah \ah}$. It is important to note from Eq.~\ref{g2Dot} that the rate of change in $g_2(t,t)$ is directly proportional to the applied field $P(t)$.

\begin{figure}[h]
\includegraphics[width=1\columnwidth]{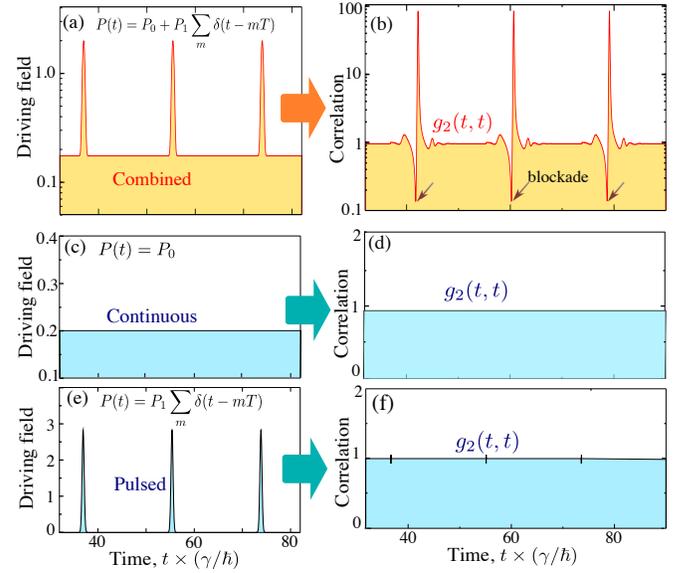}
\caption{Different driving field configurations (left panels) and the corresponding equal time correlation functions $g_2(t,t)$ (right panels). (a) schematic plot of the combined driving field that comprises the continuous and pulsed excitations. (b) the corresponding correlation function $g_2(t,t)$ as a function of time $t$ showing the strong dynamical antibunching. However, when a continuous driving field is applied alone (c), there is only conventional blockade (d) and when a short pulse is applied alone (e), the antibunching is washed out. This is because short pulses are broad in energy so the conventional blockade, which depends on energy shift of a multiple particle state out of resonance, no longer operates efficiently.
The chosen parameters are $\alpha/\gamma=0.05$, $P_0/\gamma=0.2$, $P_1/\gamma=1$, $T\gamma/\hbar=18.5$ and $E/\gamma=2$.}
\label{PulseG2}
\end{figure}

\textbf{The blockade mechanism:--} Under a conventional continuous (time independent) driving field, the system reaches its steady state where $\dot{g}_2(t,t)=0$ implying $f(t)=0$ through Eq.~\ref{g2Dot}. In such a continuous driving field configuration, the system shows the conventional blockade with a constant correlation function $g_2(t,t)=g_0$. Here we consider a driving field configuration, 
\begin{eqnarray}
P(t) = P_0 + P_1 \sum_m \delta(t-mT)
\end{eqnarray}
that is, a combination of a continuous driving field $P_0 $ and a series of $\delta$-function pulses, where $m$ is an integer. We choose the time delay between consecutive pulses $T\gamma/\hbar \gg 1$ such that the system reaches the steady state in between the pulses. Let us consider the dynamics before and after the $m$-th pulse. Just before the arrival of the pulse $(m-1)T \ll t <mT$, the system would have forgotten the effect of the previous pulse and would reach the conventional steady state $g_2(t,t)=g_0$. Immediately after the $m$-th pulse, the system moves away from the steady state due to the sudden excitation provided by the pulse. The corresponding correlation function:
\begin{eqnarray}
g_2(t,t) = g_0 + \frac{4 P_0}{\hbar}\int_0^{t-mT} f(mT+t') dt' 
\label{g2_AfterPulse}
\end{eqnarray}
where $mT<t<(m+1)T$. Importantly, the change in the correlation function, represented by the integral in Eq.~\ref{g2_AfterPulse}, is proportional to the continuous part of the driving field $P_0$. A change in $g_2(t,t)$ from $g_0$ requires both $P_0 \ne 0$ and $P_1 \ne 0$. The need of $P_0 \ne 0$ is explicit in Eq.~\ref{g2_AfterPulse}. Additionally, $P_1\ne 0$ is needed, because the change in $g_2(t,t)$ is given by the integral of $f(t)$ that can contribute only when it moves away from the steady state $f(t)=0$.  Thus, the change in $g_2(t,t)$ from its conventional (blockade) value requires the combined form of the driving field that combines the pulses with continuous excitation. Each of them individually would induce no change in the correlation function and thus the photon statistics would remain the same as that of the conventional blockade (see Fig.~\ref{PulseG2}). We emphasise that even the $\delta$-pulses, which are dynamical in nature, provide just a constant $g_2(t,t)$ in absence of the continuous excitation.

Note that the correlation function $g_2(t,t)$ would reach the steady state before the arrival of the next $(m+1)$ pulse. Thus the integral in Eq.~\ref{g2_AfterPulse} vanishes to satisfy $g_2(t,t)=g_0$ at $t=(m+1)T-\epsilon$ where $\epsilon$ is small. The total integral can be seen as a sum of contributions coming from the different time segments of the total interval from $t=mT$ to $(m+1)T-\epsilon$. Contributions from the individual segments oscillate between negative and positive values such that all contributions added together give $\int_0^{T-\epsilon} f(mT+t') dt' = 0$. Thus the system goes through the cycles of bunching (large $g_2(t,t)$) and antibunching (small $g_2(t,t)$) over time, as evident in Fig.~\ref{PulseG2}(b). For the time segments when the integral is negative, the value of the correlation function $g_2(t,t)$ can be lower than the conventional value $g_0$ and can induce stronger antibunching than the conventional one. 

\begin{figure}[h]
\includegraphics[width=1\columnwidth]{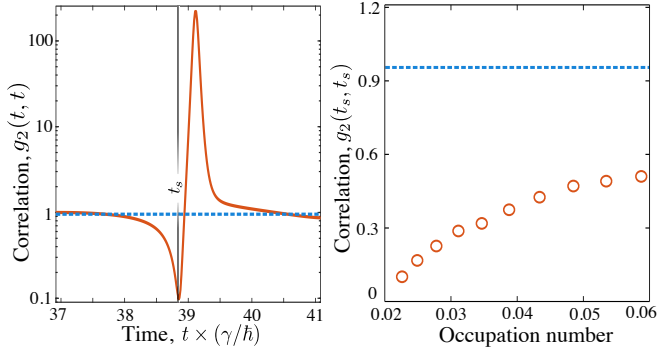}
\caption{Strong antibunching in a weakly nonlinear mode ($\alpha/\gamma\ll 1$). Left: $g_2(t,t)$ is plotted as a function of time $t$ with a combined driving field (red solid line) and with a continuous driving field (blue dotted line). We see that while the conventional $g_2(t,t)$ for a continuous driving field stays constant around $1$, the combined driving field periodically induces a small $g_2(t,t)$. Right: The correlation function $g_2(t_s,t_s)$ ( $t_s$ is indicated in the left panel) is plotted for different occupation numbers $n$ (by varying $P_0$). The calculated $g_2\approx 1$ for the conventional continuous driving field (blue dotted line). For the combined driving field, $g_2(t,t)$ (red circles) is small for all considered $n$. We used the parameters $E/\gamma=2$, $\alpha/\gamma = 0.05$, $P_0/\gamma =0.5$, $P_1/\gamma=0.5$ and $T\gamma/\hbar=18.5$ (such that the pulse arrives in the left-hand plot at $t=37\hbar/\gamma$, corresponding to the beginning of the plot scale).}

\label{WeakNonlinearity}
\end{figure}

\textbf{Analysis of equal time correlations:-} In Fig.~\ref{WeakNonlinearity}, we show the equal time correlation function $g_2(t,t)$ for a weakly nonlinear mode with $\alpha/\gamma=0.05$. For such a mode, the conventional blockade can be induced by a continuous driving field and provides a very weak antibunching with $g_2 \approx 1$. For the combined driving field, the mode shows strong antibunching in certain periodic intervals. In the combined driving field configuration, the pulses periodically  excite the mode on top of the continuous excitation. The time interval where the correlation function $g_2(t,t)$ is small follows this periodicity of the combined driving field. In the left panel of Fig.~\ref{WeakNonlinearity}, we show one such period of the correlation function $g_2(t,t)$. In the right panel, we show $g_2(t,t)$ for different occupation numbers at a time where $g_2(t,t)$ is minimum. We find that $g_2(t,t)$ remains small for the considered small occupation numbers. We are unable to find small $g_2(t,t)$ for large occupation number $n\sim 1$ in the present weakly nonlinear regime. These results are comparable to what one gets from the interference induced unconventional blockade in two-mode configuration \cite{Liew10}. However, unlike the unconventional blockade, here the unequal time correlation function does not show rapid oscillations (to be shown later in this article).

\begin{figure}[h]
\includegraphics[width=1\columnwidth]{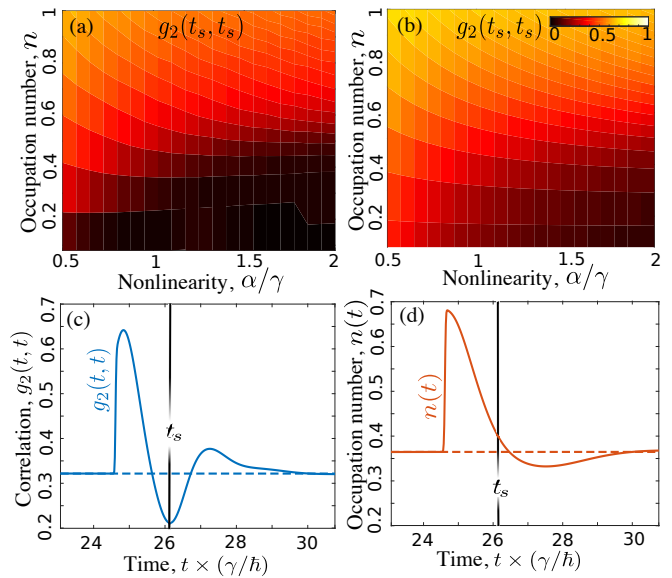}
\caption{Enhancing antibunching in the strongly nonlinear regime. Panels (a) and (b) show color plots of the correlation function $g_2(t_s,t_s)$ calculated at time $t=t_s$ (indicated in panels (c) and (d)) as functions of occupation numbers $n(t_s)$ and the nonlinearity strength $\alpha/\gamma$ for combined and continuous driving fields, respectively. We see that for a given value of $\alpha/\gamma$, the combined driving field provides a much smaller $g_2(t,t)$ compared to a continuous driving field for the same occupation number. Here we varied $P_0$ to achieve different occupation numbers. In (c) and (d), we consider the nonlinearity $\alpha/\gamma = 1$ and show the correlation function $g_2(t,t)$ and occupation number $n(t)$, respectively, as functions of time $t$ after a $\delta$-pulse is applied at $t=2T$. The dotted lines indicate the respective quantities when a constant driving field is applied (conventional blockade). We see enhancement of the single photon statistics with smaller $g_2(t,t)$ and larger $n(t)$ for certain times compared to the same for conventional blockade. We used the parameters $E/\gamma=0.25$, $P_0/\gamma=0.5$, $P_1/\gamma=0.2$ and $T\gamma/\hbar=12.3$.}
\label{StrongNonlinearity}
\end{figure}

Our considered mechanism also allows us to operate in the strongly nonlinear regime where large occupation numbers are accessible keeping $g_2(t,t)$ small. A constant driving field $P(t) = P_0$ induces conventional blockade in the strongly nonlinear regime. Under this constant driving field the system reaches to the steady state with constant $g_2(t,t)$ and $n(t)$. With a suitable $E/\gamma=0.25$ we minimize $g_2(t,t)$ for the given driving field. Keeping the same set of parameters, we introduce the additional series of $\delta$-pulses. We immediately find that the combined driving field, that is instigated by adding the pulses, induces stronger antibunching than the same for the constant driving field. Moreover, at time $t=t_s$ when the correlation $g_2(t,t)$ is minimum, the occupation number $n(t_s)$ is higher than what we get from the conventional blockade (constant driving field). Thus, the enhancement of single photon statistics under a combined driving field is two fold: a reduction in the correlation function $g_2(t,t)$ and a simultaneous increase in the mode occupation number $n$. In Fig.~\ref{StrongNonlinearity}, we show the single photon statistics of a strongly nonlinear mode. We present the color plots for the correlation function $g_2(t,t)$, obtained at time when it is minimum, induced dynamically by the combined driving field and conventionally by a continuous driving field, respectively, as functions of the nonlinear interaction strength $\alpha$ and mode occupation $n$. In the plots, the darker regions are indicating small $g_2(t,t)$ regimes. We find that the darker region for the dynamically induced blockade is larger than that of the conventional blockade. In addition, as shown in the supplemental material, the enhanced antibunching is not sensitive to our choice of $\delta$-function pulses and also appears with finite duration pulses provided they are shorter than the lifetime set by the inverse of the system decay rate.

\begin{figure}[h]
\includegraphics[width=1\columnwidth]{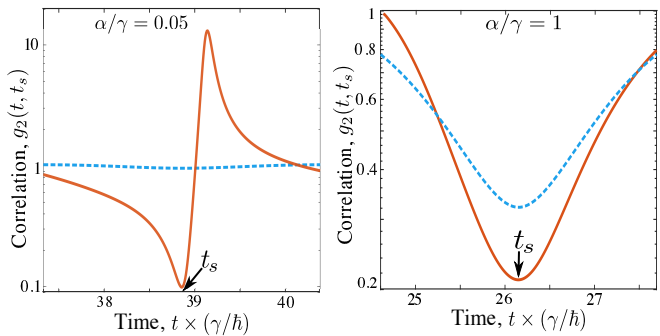}
\caption{Unequal time correlation function in different regimes of nonlinearity. While the red solid lines are representing the unequal time correlation function for the combined driving field, the blue dotted lines are showing the values corresponding to the conventional blockade. Left panel: we plot the unequal time correlation function $g_2(t,t_s)$ for $\alpha/\gamma=0.05$, where $t_s$ is a reference time as indicated in the figure. Other parameters are $E/\gamma=2$, $\alpha/\gamma = 0.05$, $P_0/\gamma =0.5$, $P_1/\gamma=0.5$ and $T\gamma/\hbar=18.5$. Right panel: $g_2(t,t_s)$ in the strong nonlinearity regime with $\alpha/\gamma=1$. Other parameters are $E/\gamma=0.25$, $P_0/\gamma=0.5$, $P_1/\gamma=0.2$ and $T\gamma/\hbar=12.3$. }
\label{UnequalTimeG2}
\end{figure}

\textbf{Analysis of unequal time correlations:-} In Fig.~\ref{UnequalTimeG2}, we show the unequal time correlation function in the weak and strong nonlinearity regimes. Our system is dynamical in nature and thus the unequal time correlation function $g_2(t,t')$ depends individually on $t$ and $t'$. We consider that the reference time $t'=t_s$, at which the equal time correlation is minimum, and evaluate the correlation function $g_2(t,t_s)$ as a function of time $t$. From the figure, we find that $g_2(t,t_s)$ remains small for $|t-t'| \sim \hbar/\gamma $ in both regimes of nonlinearity. This means that no extraordinary time resolution is needed to probe the antibunching effect in both the weak and strong nonlinearity regimes.

We compare the dynamical blockade with the conventional blockade induced by a constant driving field. In the weakly nonlinear regime, we find no sign of antibunching with $g_2(t,t_s)\approx 1$ at all times for the conventional blockade. In the strongly nonlinear regime, the conventional blockade does show an antibunching effect, with, however, a larger $g_2(t,t_s)$ compared to the same for the dynamical blockade in the most relevant regime $t\approx  t_s$. 

In the weakly nonlinear regime, the dynamical blockade is most comparable with the unconventional blockade occurring between two strongly coupled modes \cite{Liew10}. Indeed, it offers small equal time correlations similar to what we have obtained for the dynamical blockade. However, the unequal time correlation function for the unconventional blockade is controlled by a timescale that is inversely proportional to the mode coupling \cite{Bamba11}. In the required strong coupling regime, this correlation function rapidly oscillates in time. Observing the unconventional blockade thus requires high time resolution \cite{Vaneph18,Snijders18}. In our dynamical blockade, the timescale controlling $g_2(t,t_s)$ is given by the photon life time $\hbar/\gamma$, that is, a natural time resolution in emission from the mode.

\textbf{Conclusions:--} We have introduced a dynamically induced blockade mechanism that is universal in all regimes of nonlinearity strength. We have presented advantages of this dynamical blockade over the existing blockade mechanisms, conventional and unconventional. However, unlike existing blockades, the dynamical blockade is not a continuous property of the system; instead it goes through the cycles of bunching and antibunching effects over time. Strong antibunching forms only in certain periodic time windows at particular time delays from an applied pulse. To select only these time windows and to exclude all other time segments, additional arrangements in experimental setups are required. For instance, single photons can be obtained by introducing a shutter in the emission and opening it up only during the time windows when the blockade is the strongest. The required timescale of these windows is set by the inverse of the system dissipation rate. 

Dynamical blockade can be implemented in a number of systems containing nonlinear bosonic modes, e.g., optical cavities coupled to various systems~\cite{TannoudjiBook, Birnbaum05, Dayan08,Lang11,Ningyuan18, Faraon08, Lemonde16, Rabl11}, photonic crystal cavities~\cite{Akahane03} and nonlinear cavities \cite{Walmsley15}. Exciton-polaritons in semiconductor microcavities offer yet another alternative system. In fact, this could be an ideal system for exploring the dynamical blockade in both weakly and strongly interacting regimes~\cite{Kasprzak07, Sun17, Rosenberg18, Togan18}.

\section{Acknowledgements} 
This work was supported by the Singapore Ministry of Education, grant MOE2017-72-1-001.

\bibliography{references}

\end{document}